\documentclass[fleqn,twoside]{article}

\usepackage{espcrc2}

\usepackage{graphicx}

\usepackage[figuresright]{rotating}

\newcommand{\AmS}{{\protect\the\textfont2

  A\kern-.1667em\lower.5ex\hbox{M}\kern-.125emS}}

\title{Two-Pion Production in Proton-Proton Collisions --- Experimental Total
  Cross Sections and their Isospin Decomposition}
 
\author{T.~Skorodko\address[PIT]{Physikalisches Institut der Universit\"at
  T\"ubingen, D-72076 T\"ubingen, Germany},
M.~Bashkanov\addressmark[PIT],
 D.~Bogoslawsky\address[JINR]{Joint Institute for Nuclear Research, Dubna,
  Russia},
H.~Cal\'en\address[SL]{The Svedberg Laboratory, Uppsala, Sweden},
H.~Clement\addressmark[PIT],
E.~Doroshkevich\addressmark[PIT],
L.~Demiroers\address[HU]{Hamburg University, Hamburg, Germany},
C.~Ekstr\"om\addressmark[SL],
K.~Fransson\addressmark[SL],
L.~Gustafsson\address[UU]{Uppsala University, Uppsala,Sweden},
B.~H\"oistad\addressmark[UU],
G.~Ivanov\addressmark[JINR],
M.~Jacewicz\addressmark[UU],
E.~Jiganov\addressmark[JINR],
T.~Johansson\addressmark[UU],
O.~Khakimova\addressmark[PIT],
S.~Keleta\addressmark[UU],
I.~Koch\addressmark[UU],
F.~Kren\addressmark[PIT],
S.~Kullander\addressmark[UU],
A.~Kup\'s\'c\addressmark[SL],
P.~Marciniewski\addressmark[SL],
R.~Meier\addressmark[PIT],
B.~Morosov\addressmark[JINR],
C.~Pauly\address[FJ]{Forschungszentrum J\"ulich, Germany},
H.~Petr{\'e}n.\addressmark[UU],
Y.~Petukhov\addressmark[JINR],
A.~Povtorejko\addressmark[JINR],
R.J.M.Y.~Ruber\addressmark[SL],
K.~Sch\"onning\addressmark[UU],
W.~Scobel\addressmark[HU],
B.~Shwartz\address[BINP]{Budker Institute of Nuclear Physics, Novosibirsk,
  Russia},
J.~Stepaniak\address[SINS]{Soltan Institute of Nuclear Studies, Warsaw and
  Lodz, Poland},
P.~Th\"orngren-Engblom\addressmark[UU],
V.~Tikhomirov\addressmark[JINR],
G.J.~Wagner\addressmark[PIT],
M.~Wolke\addressmark[UU],
A.~Yamamoto\address[HEARO]{High Energy Accelerator Research Organization,
  Tsukuba, Japan},
 J.~Zabierowski\addressmark[SINS],
and
J.~Zlomanczuk\addressmark[UU]}

\begin{document}

\begin{abstract}

The two-pion production in pp-collisions has been investigated at CELSIUS in
exclusive measurements from threshold up to $T_p$ = 1.36 GeV. Total and
differential cross sections have been obtained
for the channels $pn\pi^+\pi^0$, $pp\pi^+\pi^-$, $pp\pi^0\pi^0$ and also
$nn\pi^+\pi^+$. For intermediate incident energies $T_p >$ 1 GeV, i.e. in the
region which is beyond the Roper excitation but at the onset of
$\Delta\Delta$ excitation, the total $pp\pi^0\pi^0$ cross section falls behind
theoretical predictions by as much as an order of magnitude near 1.2 GeV,
whereas the $nn\pi^+\pi^+$ cross section is a factor of five larger than
predicted. An isospin decompostion of the total cross
sections exhibits a $s$-channel-like energy dependence in the region of the
Roper excitation as well as a significant contribution of an isospin 3/2
resonance other than the $\Delta(1232)$. As possible candidates the
$\Delta(1600)$ and the $\Delta(1700)$ are discussed. 

\vspace{1pc}

\end{abstract}


\maketitle

Two-pion production in nucleon-nucleon collisions is an outstanding subject,
since it connects $\pi\pi$ dynamics with baryon and baryon-baryon degrees of
freedom. There is increasing evidence that the puzzling ABC
effect observed in double-pionic fusion 
reactions may possibly be traced back to an isoscalar resonance phenomenon as
source for the peculiar pion pair production in the $\pi\pi$ scalar-isoscalar
state \cite{bash,MB,panic,hcl}. By contrast the isovector $\pi\pi$
channel in double-pionic fusion behaves regularily, i.e. shows no ABC effect
and follows the expectations from conventional t-channel $\Delta\Delta$
calculations \cite{FK}. 

In view of the challenging explanation \cite{MB,panic,hcl} offered for the ABC
effect it is interesting to study for comparison the behavior of
$\pi\pi$ production in isoscalar, isovector and isotensor $\pi\pi$ channels in
those cases, where the two actively participating nucleons do not fuse into a
final nuclear bound system. From previous work it is known that the
near-to-threshold behavior of $pp \to pp\pi^+\pi^-$ and $pp \to pp\pi^0\pi^0$
channels is well understood as being dominated by excitation and decay of the
Roper resonance \cite{alv,WB,JP,TS,skor}. At higher proton beam energies  $T_p
>$ 1.2 GeV 
theoretical calculations \cite{alv} predict the t-channel $\Delta\Delta$
excitation to take over the dominant role. These calculations are
compared in Figs. 1 and 2 with the available total cross section data
\cite{dak,cverna,coch,brunt,shim,gat,eis,pickup,jan,AE} including the new data
from this work for the energy region from
threshold up to $T_p$ = 2.2 GeV. Beyond this energy the available sparse data  
indicate that the  cross sections essentially saturate.

As already demonstrated previously  \cite{dak,jan} the experimentally obtained
total cross sections for the various channels of the $NN\pi\pi$ system are
linked by isospin relations  \cite{dak,bys}, which 
may be used for a model-independent isospin decomposition of the total cross
sections. Such a decomposition in turn provides valuable insight into the
participating reaction mechanisms, in particular into the quest, which
resonances contribute. 

For the $pp$ incident channel we have four possible $NN\pi\pi$
exit channels:  $pp\pi^+\pi^-$, $pp\pi^0\pi^0$, $pn\pi^+\pi^0$ and 
$nn\pi^+\pi^+$. For these the following isospin relations apply:\\

$\sigma_{nn\pi^+\pi^+}$ = $\frac{3}{20}|M_{121}|^2 $\\
$~~~~\sigma_{pp\pi^0\pi^0}~~$ = $\frac{1}{60}|M_{121} - \sqrt5 M_{101}|^2
~~~~~~~~~~~~~~~~(1)$\\ 
$~~~~\sigma_{pp\pi^+\pi^-}$ = $\frac{1}{120}|M_{121} + 2\sqrt5 M_{101}|^2 +
\frac{1}{8}|M_{111}|^2 $\\
$~~~~\sigma_{pn\pi^+\pi^0}~$ = $\frac{3}{40}|M_{121}|^2 +
\frac{1}{8}|M_{111}|^2 + \frac{1}{4}|M_{011}|^2, $\\

where $M_{I_{NN}^fI_{\pi\pi}I_{NN}^i}$ denotes the reduced matrix element for
the isospin $I_{\pi\pi}$ of the pion pair and for the
isospins $I_{NN}^f$ and $I_{NN}^i$ of the nucleon pair in final and incident
channels, respectively. As we see from eq. (1), the simplest situation with
regard to isospin decomposition is given for the $nn\pi^+\pi^+$ channel, which
depends only on a single matrix element. We also note that in most cases the
matrix elements enter incoherently in the total cross sections \footnote{Note
  that in differential cross sections the isospin matrix elements in general
  enter coherently, which complicates the isospin decomposition there. Hence we
  stick here to the treatment of total cross sections and leave the discussion
  of the differential cross sections for a forthcoming paper}. Only in the
$pp\pi^0\pi^0$ and $pp\pi^+\pi^-$ channels the matrix elements $M_{121}$ and
$M_{101}$ enter coherently, i. e. lead to an interference term. Since in the
latter the relative phase $\phi$ between both matrix elements enters, we 
may rewrite the expressions for the $pp\pi^0\pi^0$ and $pp\pi^+\pi^-$ channels
in the following way: \\

$\sigma_{pp\pi^0\pi^0}$ = $\frac{1}{60} |M_{121}|^2 +
\frac{1}{12}|M_{101}|^2  -$\\
$~~~~~~~~~~~~~~~~~~- \frac{1}{\sqrt180}|M_{121}| |M_{101}| cos\phi
~~~~~~~~~~~(2)$\\ 
$~~~\sigma_{pp\pi^+\pi^-}$ = $\frac{1}{120} |M_{121}|^2 + 
\frac{1}{6}|M_{101}|^2  + \frac{1}{8} |M_{111}|^2$\\
$~~~~~~~~~~~~~~~~~~+ \frac{1}{\sqrt180}|M_{121}| |M_{101}| cos\phi. $\\

Thus for each incident energy we have the five parameters $|M_{121}|$,
$|M_{111}|$, $|M_{101}|$, $|M_{011}|$ and $\phi$ to be determined from the
four experimental values for the total cross sections of the channels
$pp\pi^+\pi^-$, $pp\pi^0\pi^0$, $pn\pi^+\pi^0$ and $nn\pi^+\pi^+$. Since this
system is underdetermined, there is {\it a priory} no unique solution and we
need additional physics input. Such a piece of input is provided by the
information that $|M_{111}|$ got to be small compared to  $|M_{101}|$ for
three reasons: 
First of all, $|M_{111}|$ must vanish in the near-threshold region, since by
Bose symmetry isovector $\pi\pi$ pairs must be associated with $p$-waves, which
vanish at threshold \footnote{Indeed, in the near-threshold region, which is
well covered by the exclusive measurements at CELSIUS and COSY, we find 
to good approximation

$~~~\sigma_{pp\pi^+\pi^-} (T_p)$ =  2~$\sigma_{pp\pi^0\pi^0} (T_p - 20
MeV)$ 

for $T_p <$ 800 MeV within experimental uncertainties, where the 20 MeV
correction of the lab beam energy $T_p$ accounts for the different reaction
thresholds due to different $\pi^{\pm}$ and $\pi^0$ masses. I.e. in this region
both cross sections are well explained by $M_{101}$ solely.}.  
Second, the
Roper excitation, which is the leading process in the near-threshold region
\cite{alv,WB,JP,TS,skor}, contributes only a tiny fraction of its strength to
$|M_{111}|$ and 
finally , the $\Delta\Delta$ process, which is the leading process at
higher energies does not contribute at all to $|M_{111}|$. From this we
conclude that $|M_{111}|$ must play a minor role and the main difference between
$\sigma_{pp\pi^0\pi^0}$ and  $\sigma_{pp\pi^+\pi^-}$ cross sections must be
associated to the interference term, which then fixes $\phi$ (see eq. (2)).

\begin{figure}
\begin{center}

\includegraphics[width=0.45\textwidth]{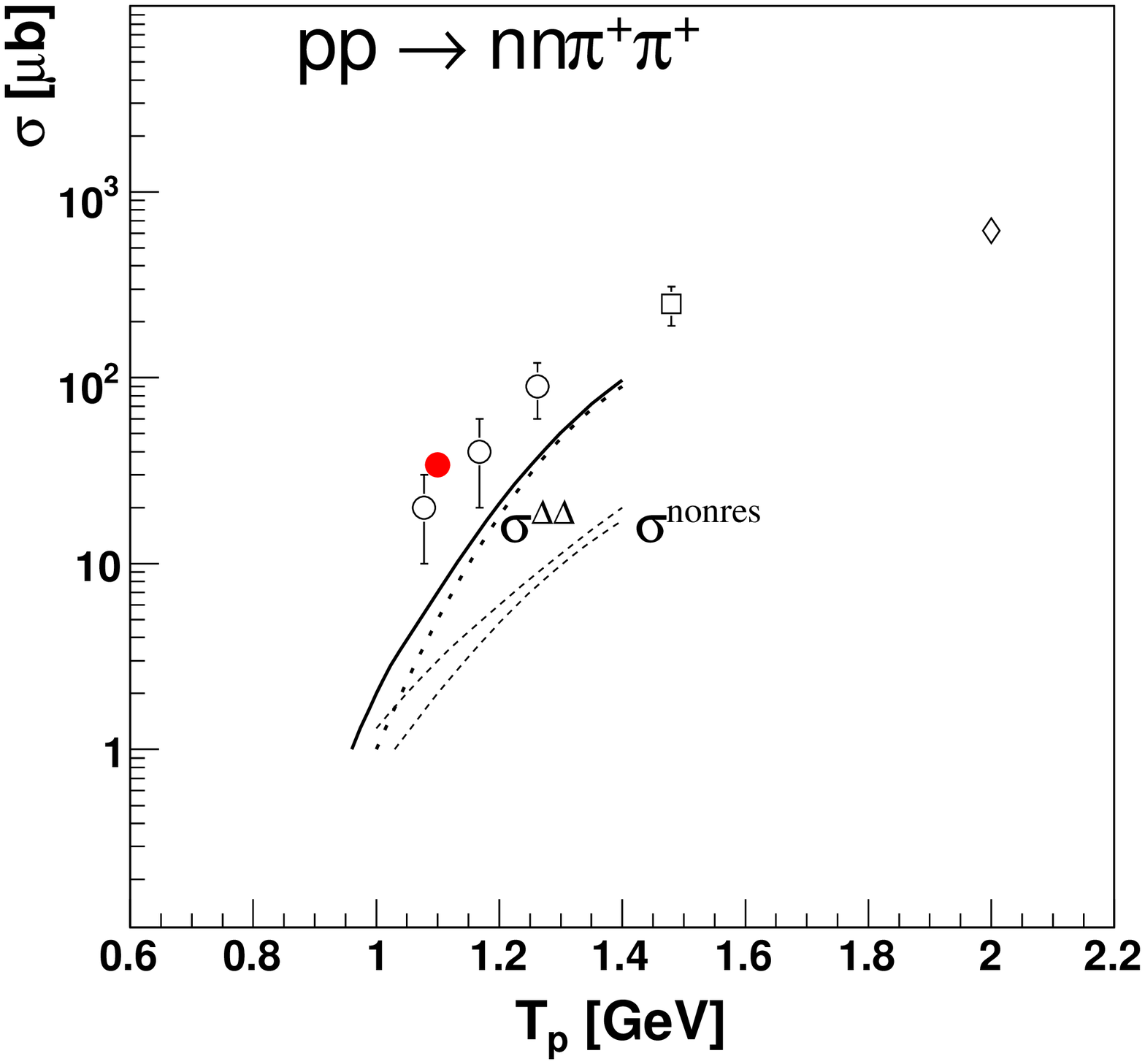}
\includegraphics[width=0.45\textwidth]{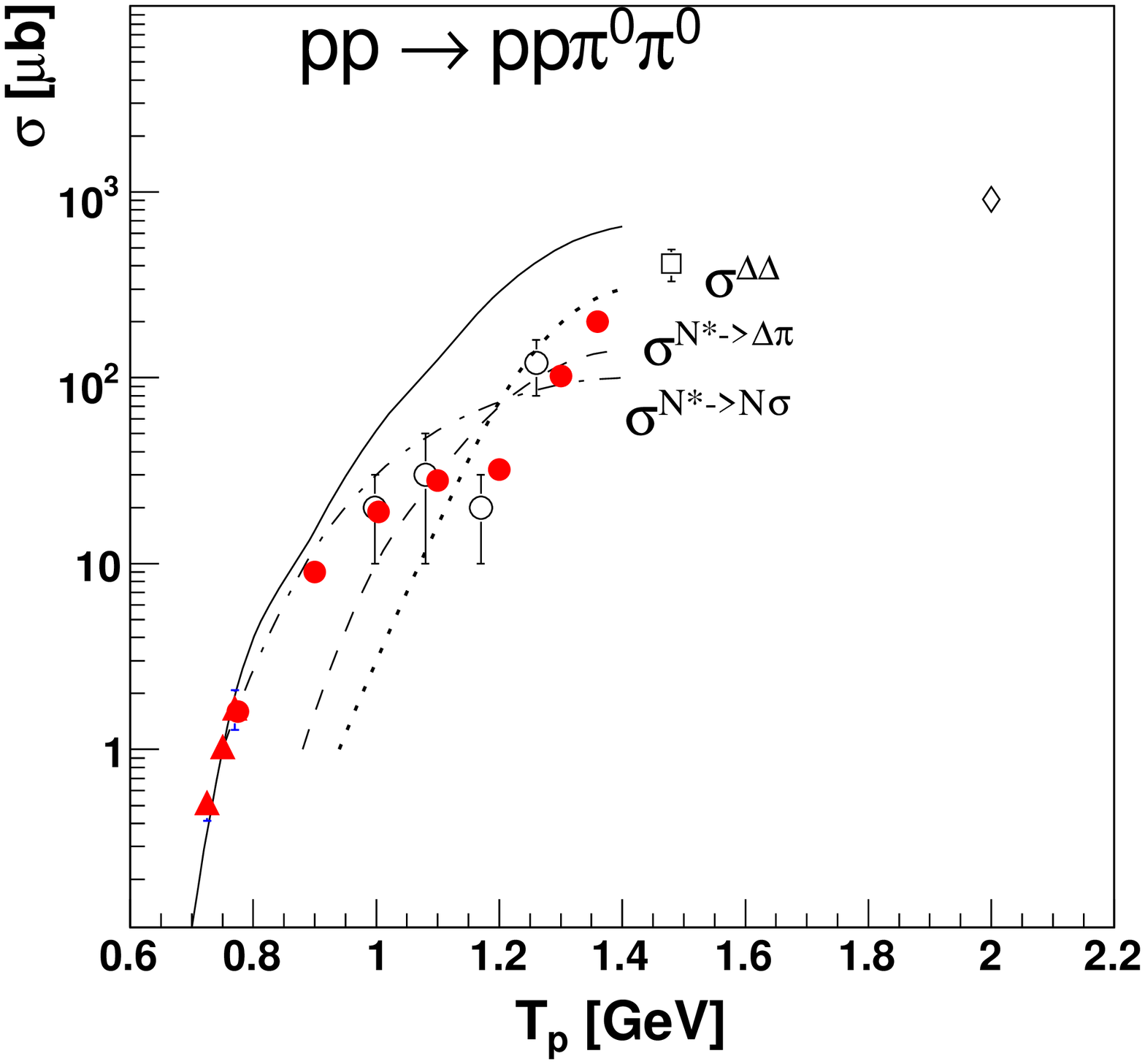}

\caption{ 
   Energy dependence of the total cross sections for the $pp \to nn\pi^+\pi^+$
   ({\bf top}) and $pp \to pp\pi^0\pi^0$ ({\bf bottom})
  reactions. Thick solid dots represent the experimental
  results of this work. The filled triangles give PROMICE/WASA results
  \cite{jan}. Open symbols denote previous bubble-chamber measurements 
  \cite{shim,eis,pickup,jan}. The drawn curves show the 
  predictions of Ref. \cite{alv} for non- and semi-resonant (short dashed:
  contributions 
  from diagrams (1) - (3) and (6), (7), respectively in Ref. \cite{alv}),
  $\Delta\Delta$ (<dotted) and Roper ($N^* \to N\sigma$: short and long
  dashed, $N^* \to \Delta\pi$: long dashed)  
  excitation processes. The solid lines denote the full calculations. 
}
\label{fig1}
\end{center}
\end{figure}

\begin{figure} 
\begin{center}
\includegraphics[width=0.45\textwidth]{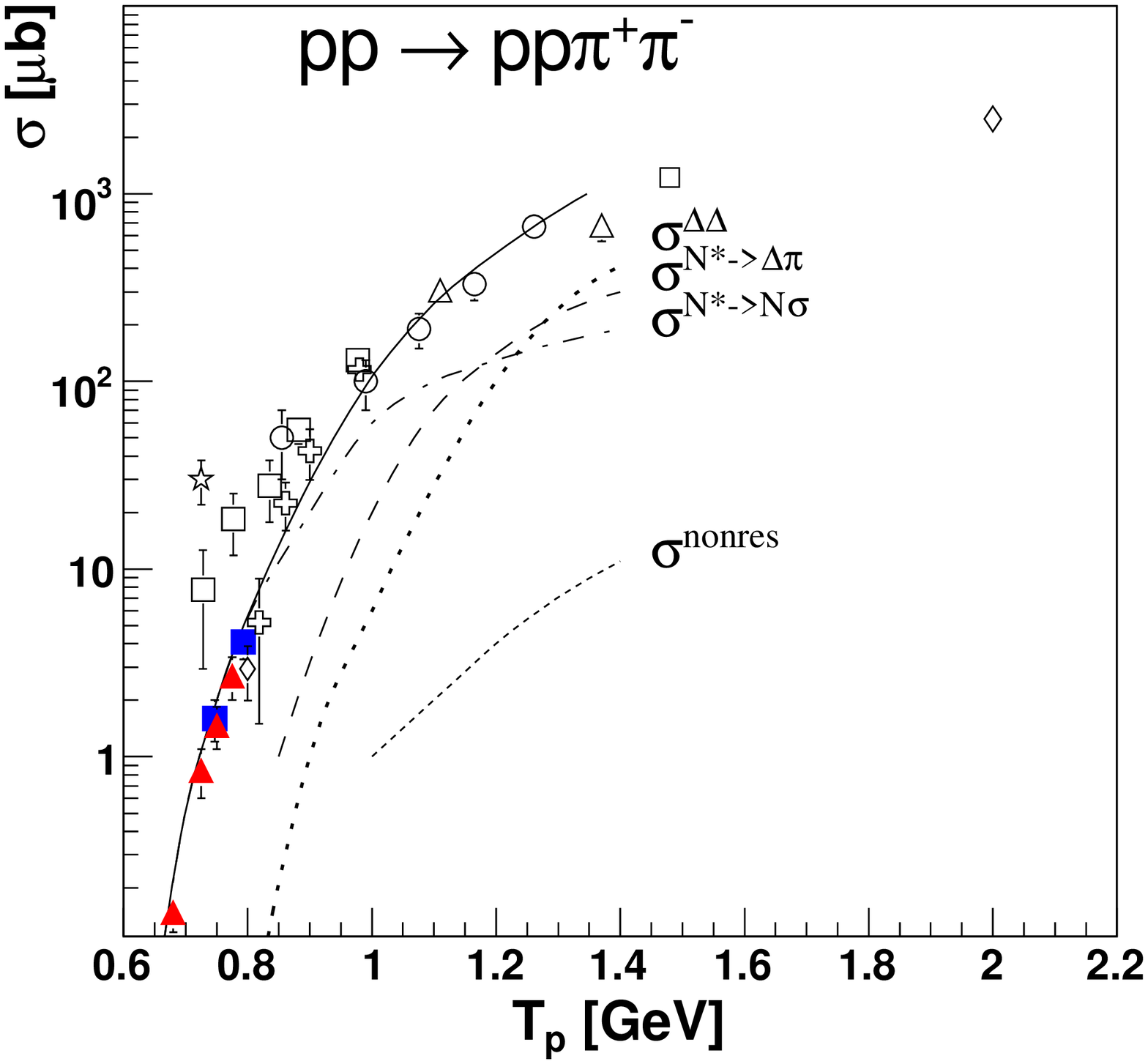}
\includegraphics[width=0.45\textwidth]{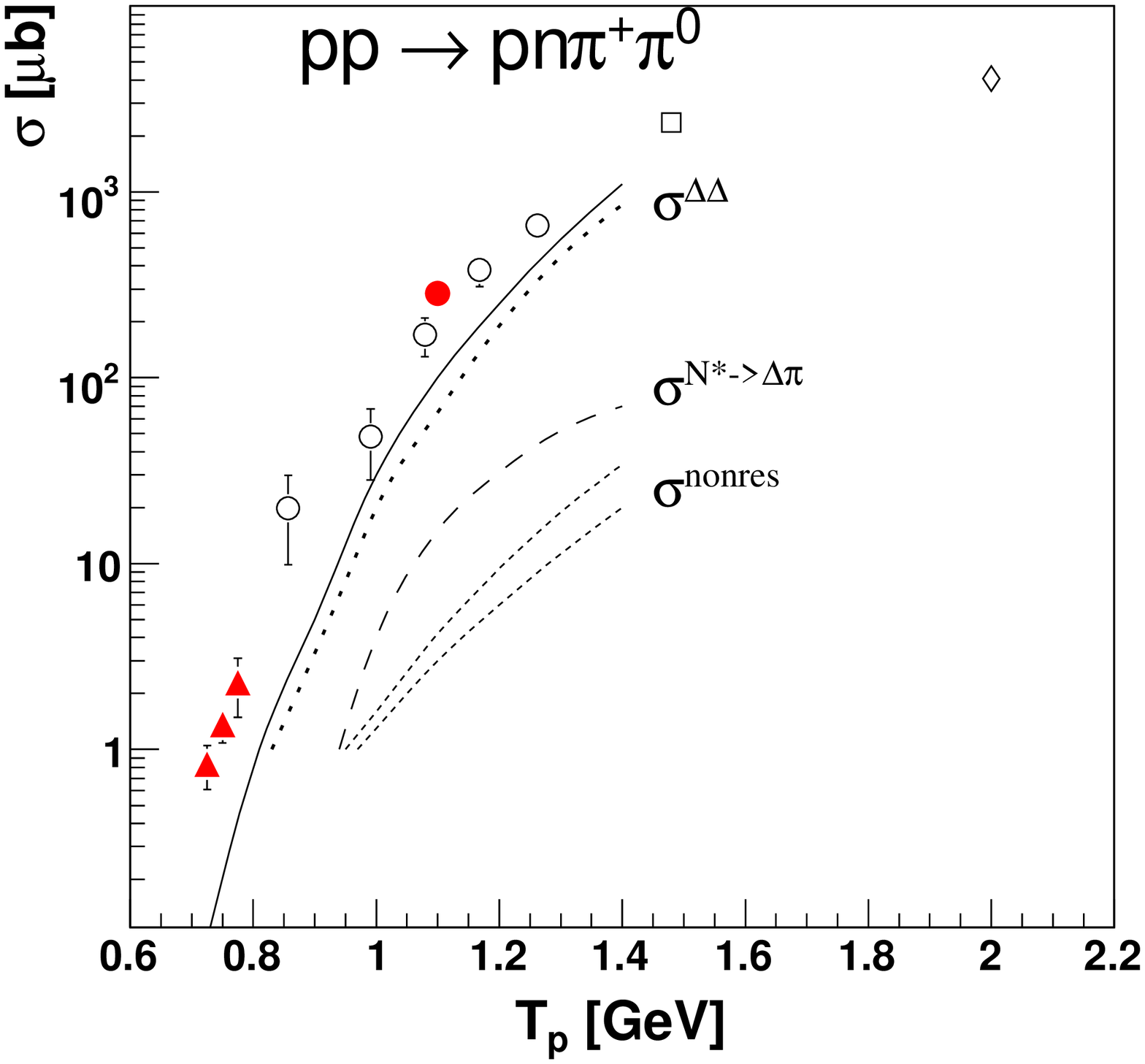}

\caption{ 
   The same as Fig. 1 but for for the $pp \to pp\pi^+\pi^-$
   ({\bf top}) and $pp \to pn\pi^+\pi^0$ ({\bf bottom})
  reactions. The data are from Refs. 
\cite{dak,cverna,coch,brunt,shim,gat,eis,pickup,jan,AE} and this work (solid
dot). The open symbols denote bubble-chamber
\cite{dak,brunt,shim,gat,eis,pickup} results and for $pp \to pp\pi^+\pi^-$
also single-arm magnetic spectrometer measurements from LAMPF
\cite{cverna,coch}. The filled triangles give PROMICE/WASA results \cite{jan},
whereas filled squares (top figure only) show COSY-TOF measurements \cite{AE}. 
}
\label{fig2}
\end{center}
\end{figure}

This suggests the following strategy:
The total cross sections of the $nn\pi^+\pi^+$ channel uniquely determines
$|M_{121}|$ at each incident energy. Having fixed the phase $\phi$ from
inspection of $pp\pi^+\pi^-$ and $pp\pi^0\pi^0$ cross sections we can
derive $|M_{101}|$ from the values of the  $pp\pi^0\pi^0$ cross section.
After having determined $|M_{111}|$ subsequently from the  $pp\pi^+\pi^-$ cross
sections, we confront our results with the expectation of resonances excited
in these reactions.

Since the available data base has been quite
restricted strongly hampering previous isospin decompostion
analyses, we have carried out a systematic program of exclusive two-pion
production measurements in $pp$ collisions from threshold up to $T_p$ = 1.36
GeV using the WASA detector  
\cite{barg} with the hydrogen pellet target system at the CELSIUS storage
ring of the Theodor Svedberg Laboratory in Uppsala. The detector has nearly
full angular coverage 
for the detection of charged and uncharged particles. The forward detector
consists of a thin window plastic scintillator hodoscope at the exit of the
scattering chamber, followed by straw tracker, plastic scintillator quirl and
range hodoscopes, whereas the central detector comprises in its inner part a
thin-walled superconducting magnet containing a minidrift chamber for tracking
and in its outer part a plastic scintillator barrel surrounded by an
electromagnetic calorimeter consisting of 1012 CsI (Na) crystals. 

Protons and neutrons have been detected in the forward detector. The protons
have been identified by the $\Delta$E-E technique,  neutrons by the condition
of having no signals in the thin detectors window hodoscope, straw tracker and
quirl. Charged pions and photons have been detected and identified in the
central detector. 

The absolute normalization of the data has been achieved by normalizing
simultaneously measured elastic scattering and/or single pion production cross
sections to known values.

Total as well as differential cross
sections have been obtained  for $pn\pi^+\pi^0$, $pp\pi^+\pi^-$,
$pp\pi^0\pi^0$ and also $nn\pi^+\pi^+$ channels. Here we concentrate on total
cross sections, which are shown in Figs. 1 - 4  together with previous
experimental results
\cite{WB,JP,dak,cverna,coch,brunt,shim,gat,eis,pickup,jan,AE} as well as to
the theoretical predictions \cite{alv}. For the $pp\pi^0\pi^0$ channel total
cross section values of 1.6(1), 9(1), 19(2), 28(3), 32(4), 102(16), 198(25)
$\mu b$ at $T_p$ = 0.775, 0.9, 1.0, 1.1, 1.2, 1.3 and 1.36 GeV have  been
obtained. For $nn\pi^+\pi^+$ and $pn\pi^+\pi^0$ channels the 
results at $T_p$ = 1.1 GeV are 34(4) and 284(30) $\mu b$, respectively. These
data are plotted in Figs. 1 - 4 by filled circles. Previous results from
PROMICE/WASA \cite{WB,JP,jan} are shown by filled triangles and those from
COSY-TOF \cite{AE} by filled squares. All these data stem from exclusive
measurements of solid statistics. The other previous data denoted by open
symbols in Figs. 1 - 4 are bubble chamber results, partly of very low
statistics, or single-arm magnetic spectrometer measurements from LAMPF
\cite{cverna,coch}. 

As we see from
Figs. 1 and 2 our measurements are in good agreement with previous experimental
results, in particular also with previous bubble chamber results
\cite{shim,eis} for the $nn\pi^+\pi^+$ and $pp\pi^0\pi^0$ channels, where we
observe strong deviations from the theoretical predictions for  $T_p >$ 1 GeV.

Two surprising features emerge from these measurements for
incident energies $T_p >$ 1 GeV, i.e. in the region, which is
beyond the Roper excitation, at the onset of the $\Delta\Delta$ excitation :
\begin{itemize}
\item Whereas the experimental $pp\pi^0\pi^0$ cross sections agree well with
  theoretical predictions \cite{alv} up to $T_p \approx$  1 GeV, i.e. in the
  region 
  of the Roper excitation, they disagree strongly at higher energies. 
  The predicted cross section keeps rising smoothly with increasing $T_p$,
  however, the
  data level off for $T_p >$ 1.0 GeV, falling behind the predictions by an
  order of magnitude near 1.2 GeV and exhibiting then a sharp rise at $T_p >$
  1.2. This behavior is also in contrast to that observed in the
  $pp\pi^+\pi^-$ channel, which is well described by the theoretical
  calculations. 
\item In sharp contrast the observed $nn\pi^+\pi^+$ cross sections are a
  factor of 5 larger than the theoretical predictions. This is very
  surprising, since  under the assumption that $\Delta\Delta$ excitation is
  the dominant process we would expect the  $pp\pi^0\pi^0$ cross section to be
  four times larger than the  $nn\pi^+\pi^+$ cross section by use of simple
  isospin arguments, since in this case both channels originate from the same
  intermediate  $\Delta^+\Delta^+$ system.
\end{itemize}

In order not to depend on model assumptions we have carried out an isospin
decomposition of the total cross sections as outlined above. 
In order to get the decomposition for all energies, which have
been measured, we draw smooth curves through
the data points and for the $nn\pi^+\pi^+$ channel we also extrapolate to lower
energies guided by the slope of the theoretical predictions (shaded curve in
Fig. 3, where the vertical extension of the shaded area indicates the assumed
uncertainty of the extracted isospin decomposed cross sections). 

We use the following convention for showing the isospin
decomposed results in Figs. 3 and~4: the results for the matrix elements are
shown as their incoherent contribution to the cross section of interest.
I.e., the result for, e.g.,  $|M_{121}|$ is shown in the plot for the
$nn\pi^+\pi^+$ channel as $\sigma_{121}^{++}$  = $\frac{3}{20}|M_{121}|^2 $ and in
the $pp\pi^0\pi^0$ channel as $\sigma_{121}^{00}$  = $\frac{1}{60}|M_{121}|^2$
etc., see shaded and dotted curves in Fig. 3 and 4, respectively.

In a first step we assume $|M_{111}|$ to be neglegible for reasons given
above. Since  $|M_{121}|$ is uniquely fixed by $\sigma_{nn\pi^+\pi^+}$, we
derive then $|M_{101}|$ and the phase $\phi$ from $\sigma_{pp\pi^0\pi^0}$ and
$\sigma_{pp\pi^+\pi^-}$ . In the near-threshold region, where these cross
sections scale in good approximation like 1:2, i.e. like the $|M_{101}|$
contributions do, and where $\sigma_{121}^{00}$ and $\sigma_{121}^{+-}$ are
lower by two orders of magnitude, the $cos\phi$ interference term does not
contribute significantly and hence $\phi$ is not well determined in this
region, However, for the region  $T_p >$ 1.0 GeV we immediately see
that the isoscalar-isotensor interference must be strongly constructive
for $\sigma_{pp\pi^+\pi^-}$  and strongly destructive for
$\sigma_{pp\pi^0\pi^0}$, in order to account for the experimental observation
that $\sigma_{pp\pi^+\pi^-}$ is larger than $\sigma_{pp\pi^0\pi^0}$ by up to an
order of magnitude. Actually we only can arrive at a reasonable description of
both channels in this region, if the interference is maximal, i.e. $\phi$~=~0.
That way we also obtain $|M_{101}|$ from the $pp\pi^0\pi^0$ cross section
plotted as $\sigma_{101}^{00}$ (shaded curve in Fig. 3). 

The cross sections  $\sigma_{101 + 121}$ calculated
from the $|M_{101}|$ and  $|M_{121}|$ contributions (including their
interference) are given in Figs. 3 and 4
by solid ($\sigma_{tot}^{00} = \sigma_{101 + 121}^{00}$) and dot-dot-dashed
($\sigma_{101 + 121}^{+-}$) 
lines, respectively. We see that in the  $pp\pi^+\pi^-$ channel not only the
near-threshold region is well reproduced, but also the high-energy
region, where  $pp\pi^0\pi^0$ and $pp\pi^+\pi^-$ data behave very differently.
Unfortunately the $pp\pi^+\pi^-$data in this energy region show quite some
scatter, which hampers severely a quantitative extraction of $|M_{111}|$. In
fact, due to the propagation of the uncertainties in the data the
uncertainties in $\sigma_{111}^{+-}$ get unpleasantly large at higher energies
(shaded area in Fig.4). We only can state
that, since $\sigma_{101 + 121}$ is low compared to the data by up to 30 $\%$
in the region 0.9 GeV $ \leq T_p \leq $ 1.2 GeV, the $|M_{111}|$ contribution
must be of this order of magnitude there. A better fixation of$|M_{111}|$
definitely needs high-quality $\sigma_{pp\pi^+\pi^-}$ data, which are not
available at present.   

Since continuing with the determination of 
$|M_{011}|$ that way would mean even larger uncertainties for this matrix
element, we instead confront our results with the expectation from a
resonance scenario based on the isospin decomposition results for $|M_{101}|$
and  $|M_{121}|$ .

\begin{figure}
\begin{center}
\includegraphics[width=0.45\textwidth]{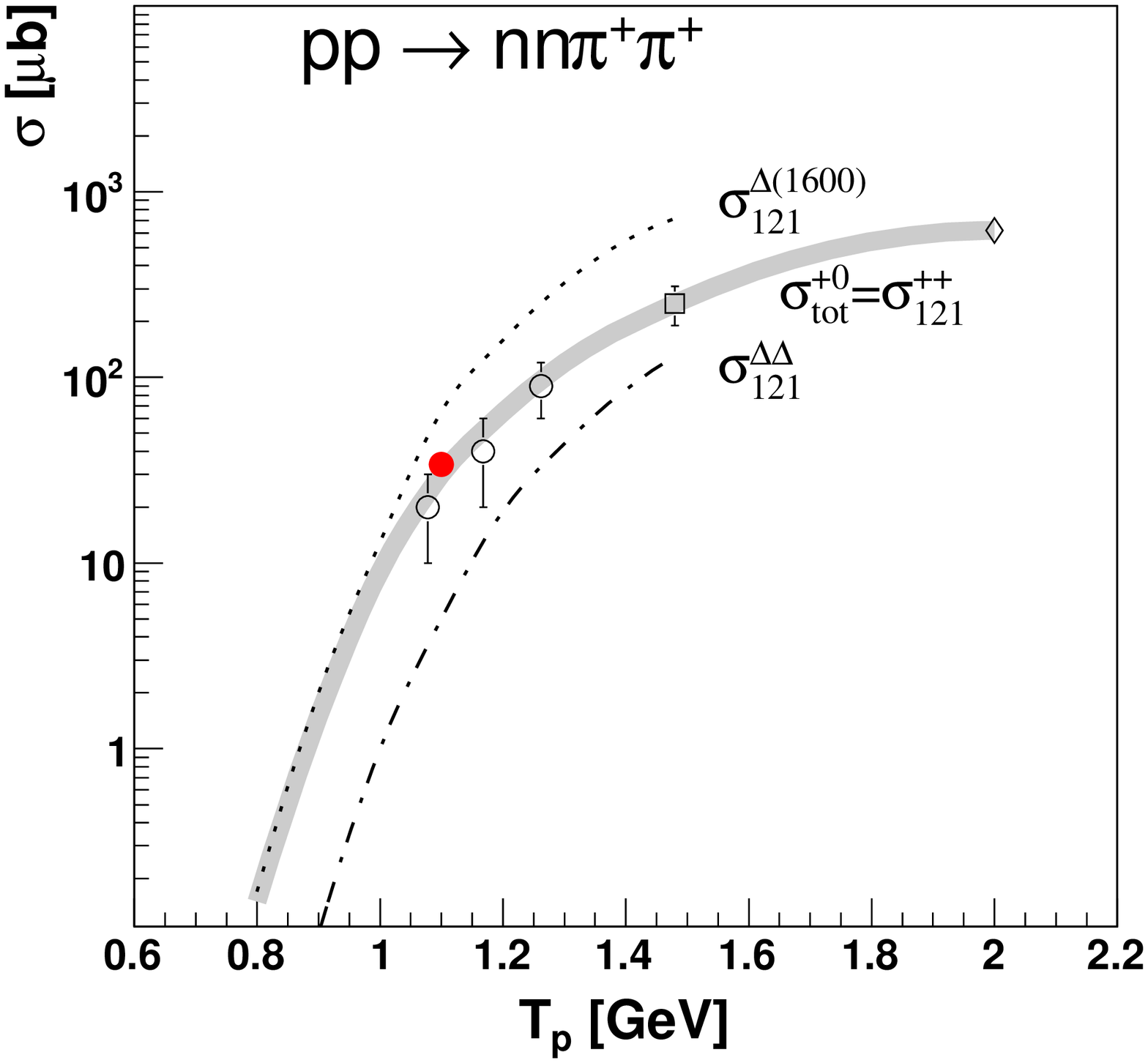}
\includegraphics[width=0.45\textwidth]{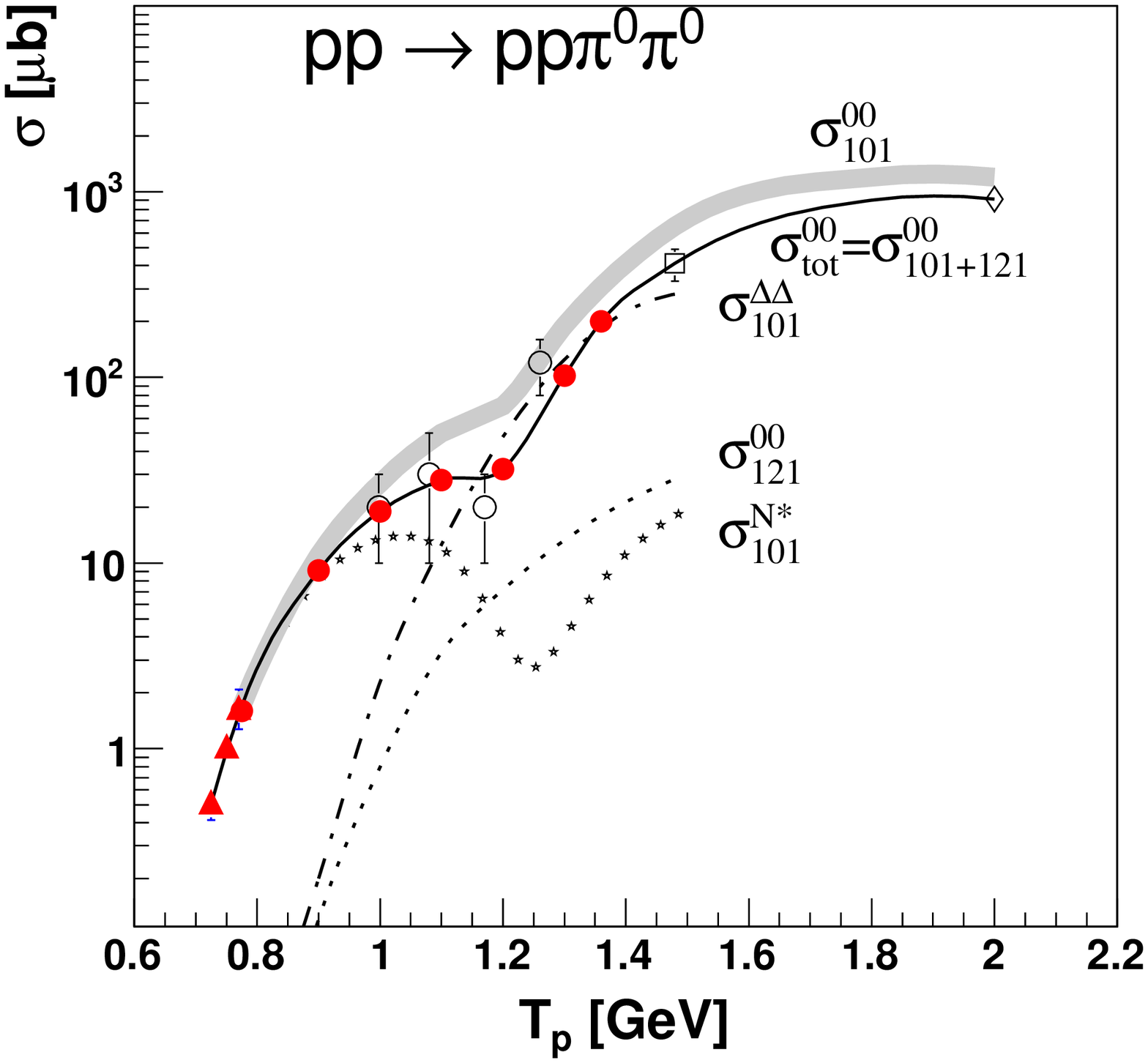}
\caption{ 
   Isospin decompostion of the total cross sections for the $pp \to
  nn\pi^+\pi^+$ ({\bf top}) and $pp \to pp\pi^0\pi^0$ ({\bf bottom})
  reactions. Data are represented by open and filled symbols, see caption of
  Figs. 1 - 2. The drawn lines show the extracted contributions of
  $\sigma_{121}$, $\sigma_{101}$ and $\sigma_{101 + 121}$ as indicated in the
  figures. The decomposition of $\sigma_{121}$ into $\Delta(1600))$ and
  $\Delta\Delta$ contributions as well as 
  of  $\sigma_{101}$ into $N^*$ and $\Delta\Delta$ contributions are also
  shown. All these curves are estimated to have uncertainties of the same order
  as indicated by the shaded area displayed for  $\sigma_{121}^{++}$ (top) and
  $\sigma_{101}^{00}$ (bottom).
}
\label{fig3}
\end{center}
\end{figure}

\begin{figure}
\begin{center}
\includegraphics[width=0.45\textwidth]{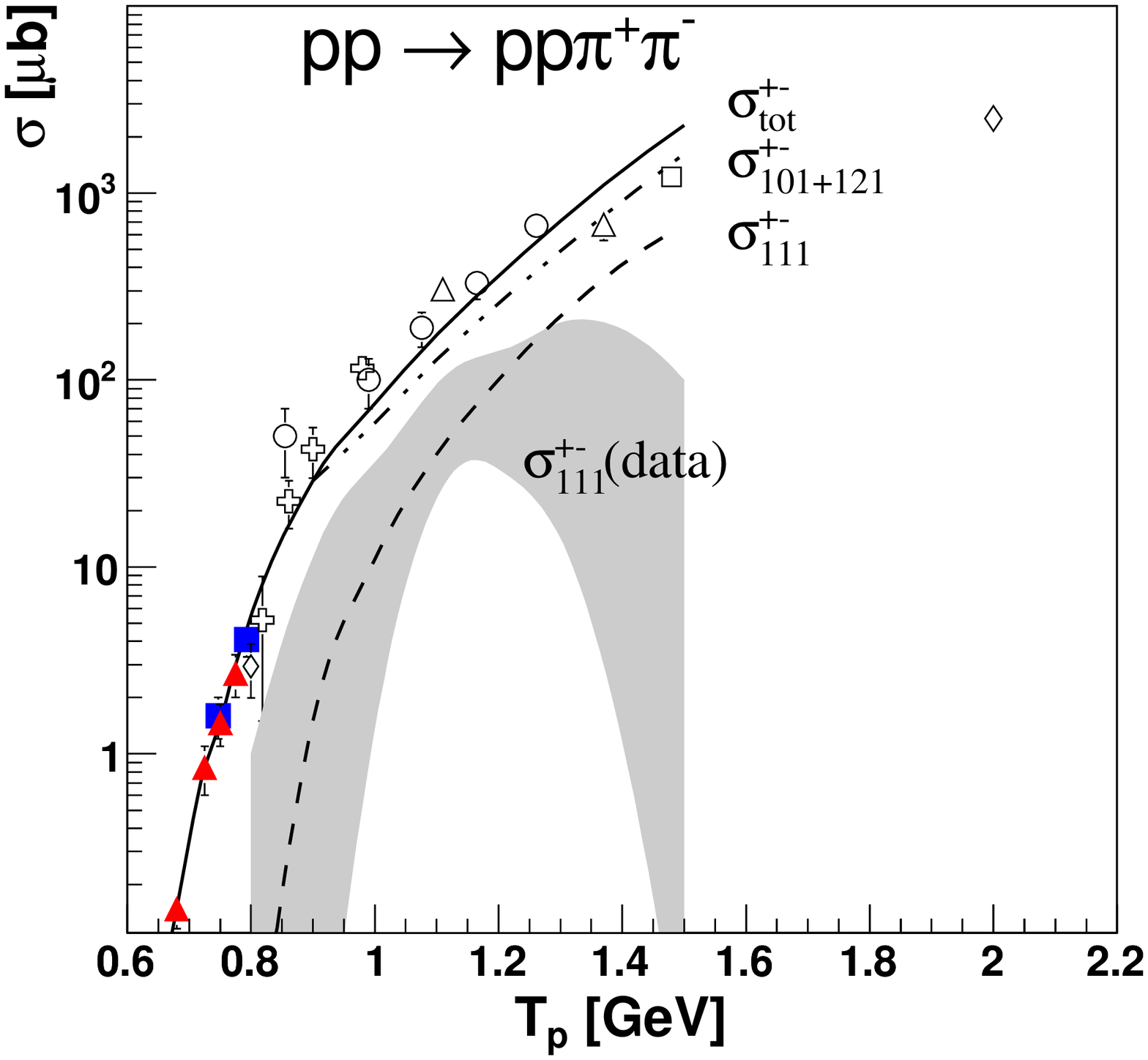}
\includegraphics[width=0.45\textwidth]{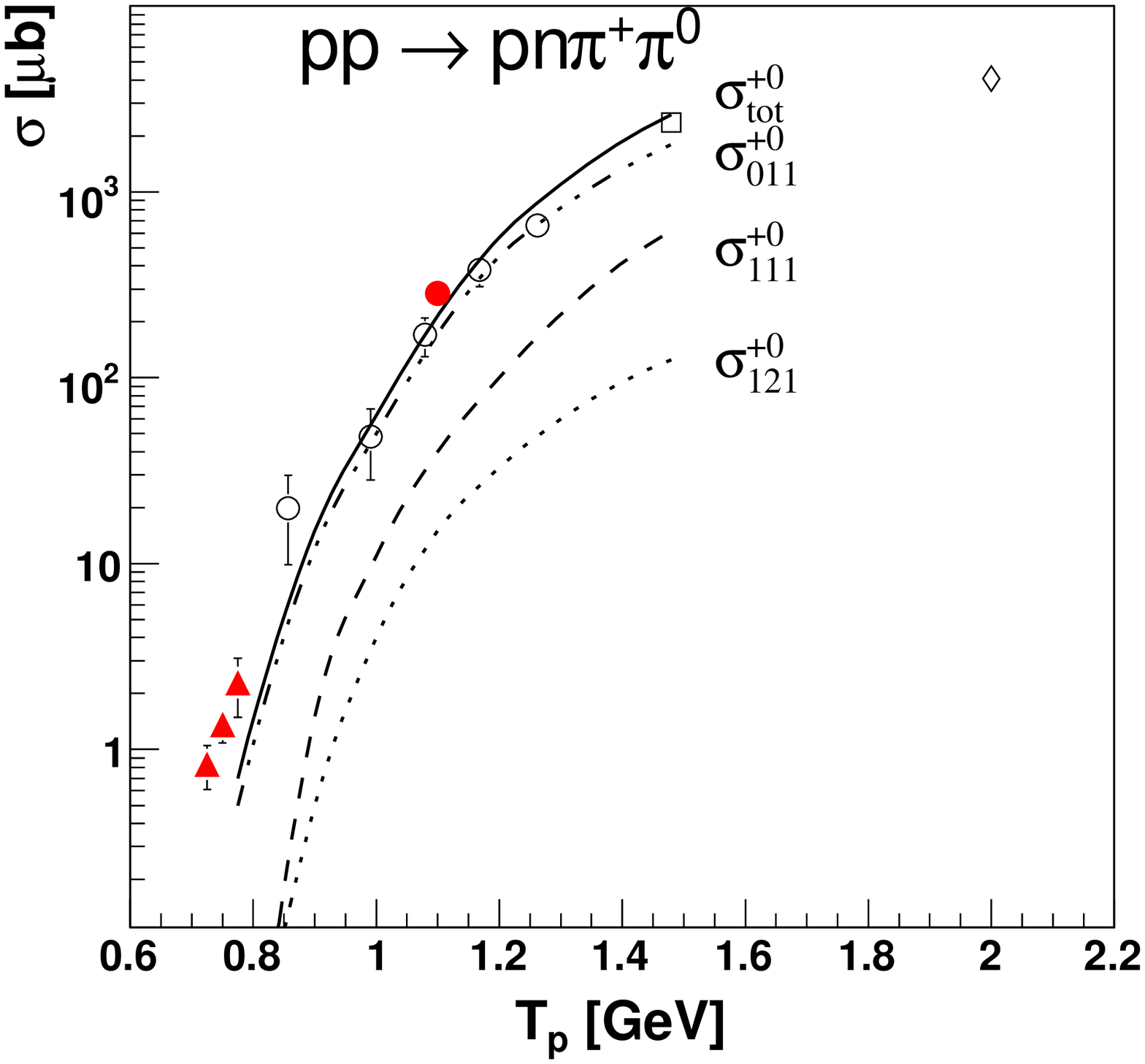}

\caption{ 
   Same as Fig. 3, but for the $pp \to
  pp\pi^+\pi^-$ ({\bf top}) and $pp \to pn\pi^+\pi^0$ ({\bf bottom})
  reactions. Data are represented by open and filled symbols, see caption of
  Figs. 1 - 2. Broken lines show the extracted contributions of
  $\sigma_{121}$, $\sigma_{111}$, $\sigma_{011}$ and $\sigma_{101 + 121}$ as
  indicated in the figures, whereas solid lines denote their sum providing the
  total cross sections for $pp \to pp\pi^+\pi^-$ ($\sigma_{tot}^{+-} =
  \sigma_{101+121}^{+-} + \sigma_{111}^{+-}$) and $pp \to pn\pi^+\pi^0$
  ($\sigma_{tot}^{+0} = \sigma_{011}^{+0} + \sigma_{111}^{+0 } +
  \sigma_{121}^{+0}$). The large shaded area indicates
  $\sigma_{111}(data)$ as derived by direct comparison of the $pp \to
  pp\pi^+\pi^-$ reaction data with  $\sigma_{101 + 121}$, whereas
  $\sigma_{111}$ (dashed line) is derived by use of relations (3) - (6).
}
\label{fig4}
\end{center}
\end{figure}

In the resonance scenario we have the excitation of one or two resonances in
the intermediate step of the reaction process, which then decay by pion
emission into the $NN\pi\pi$ channels. In particular we have the excitation of
the Roper or other higher-lying $N^*$ resonances schematically described by
$pp \to NN^* \to NN\pi\pi$, the $\Delta\Delta$ excitation ($pp \to
\Delta\Delta \to NN\pi\pi$ ) and -- as we shall see below -- the excitation of
the $\Delta(1600)$ ($pp \to N\Delta(1600) \to N\Delta\pi \to NN\pi\pi$) or
possibly also $\Delta(1700)$.

For each particular resonance scenario the reduced isospin
matrix elements are linked by simple isospin relations obtained from isospin
recoupling by use of $9j$ symbols. For $N^*$ excitation we obtain for the
decay branch $N^* \to N\sigma$ : \\

$M_{011}^{N^* \to N\sigma} = M_{111}^{N^* \to N\sigma} = M_{121}^{N^* \to
  N\sigma} =
0~~~~~(3)$\\  

and for the decay branch $N^* \to \Delta\pi$ :\\

$M_{111}^{N^* \to \Delta\pi} = + \frac{1}{2} M_{101}^{N^* \to
  \Delta\pi}$\\
$~~~~~M_{011}^{N^* \to \Delta\pi} = + \sqrt \frac{1}{2} M_{101}^{N^* \to
  \Delta\pi}~~~~~~~~~~~~~~~~~~~~~~~~~~~~(4)$\\
$~~~~M_{121}^{N^* \to \Delta\pi} = 0$\\

The relation between both branches of the Roper process is well-known from the
studies of the $pp \to pp\pi^+\pi^-$ and $pp \to pp\pi^0\pi^0$ reactions in
the near-threshold region \cite{alv,WB,JP,TS,skor}.  
For the $\Delta\Delta$ excitation we get: \\

$M_{101}^{\Delta\Delta} = - \sqrt5 M_{121}^{\Delta\Delta}$\\
$~~~~M_{011}^{\Delta\Delta} = + \sqrt \frac{15}{2}
M_{121}^{\Delta\Delta}~~~~~~~~~~~~~~~~~~~~~~~~~~~~~~~~~(5)$\\ 
$~~~~M_{111}^{\Delta\Delta} = 0$  \\

and for the excitation of the $\Delta(1600)$ and its dominant decay branch
$\Delta(1600) \to \Delta\pi \to N\pi\pi$ we find: \\

$M_{111}^{\Delta(1600)} = + \sqrt \frac{5}{3} M_{121}^{\Delta(1600)}$\\
$~~~~M_{011}^{\Delta(1600)} = + \sqrt \frac{10}{3}
M_{121}^{\Delta(1600)}~~~~~~~~~~~~~~~~~~~~~~~~~~~~~~~~~(6)$\\ 
$~~~~M_{101}^{\Delta(1600)} = 0.$  \\

From this we see that the two-pion decay of $N^*$ resonances does not
contribute to $\sigma_{121}$ and hence also not to  $\sigma_{pp \to
  nn\pi^+\pi^+}$. Therefore this cross section should be completely covered by
the 
$\Delta\Delta$ process aside from small non-resonant contributions. As 
already discussed above, we face the problem that the theoretical calculations,
which take into account both these processes, severely underpredict the
measured data for this channel by a factor of
four, which naively would mean that the $\Delta\Delta$ process ought to be
larger by that factor. However, such an enlargement of this process would
severely destroy any understanding of the data at high energies in the
other channels as we may easily judge from inspection of Figs. 1 and 2. Hence
we conclude that another resonance mechanism must be present, which
strongly contributes to the $nn\pi^+\pi^+$ channel and $\sigma_{121}$,
respectively, but not to $\sigma_{101}$. Since only the excitation of I~=~3/2
resonances and their successive decay into $N\pi\pi$ can fulfill these
conditions, we have to look for a higher lying $\Delta$ excitation. The next
higher-lying candidate is $\Delta(1600)$, which actually due to its
large width of about 350 MeV appears not unlikely to contribute 
already at the energies of interest. Since $\Delta(1600)$ preferably decays via
$\Delta(1232)$, it contributes strongly to the isotensor cross section and
also to the isovector part, but not to the isoscalar one. Similar arguments
hold also for the $\Delta(1700)$, which has the disadvantage of having the
pole at still higher mass, but has the advantage of a $s$-wave decay
$\Delta(1700) \to \Delta \pi$.

With this hypothesis and the constraint $\phi = 0$ we reconstruct the
amplitude for the $\Delta(1600)$ excitation from
$\sigma_{nn\pi^+\pi^+}$ assuming that the theoretical prediction for the
$\Delta\Delta$ process must be essentially correct both in its magnitude and
in its energy dependence. The latter should be particularily reliable, since
its involves the well-known $\Delta$ propagators with double $p$-wave pion
emission~\footnote{However, we do not
  dare to extrapolate the behavior of the $\Delta\Delta$ cross section up to 2
GeV. Hence we stop the decompositions in partial cross sections at 1.4 GeV.}.

From eqs. (5) we see that $M_{101}^{\Delta\Delta}$ and
$M_{121}^{\Delta\Delta}$ are of opposite sign, but the condition $\phi$ =
0 requires the complete matrix elements $M_{101}$ and $M_{121}$ to be of the
same sign. Hence $\Delta\Delta$ and $\Delta(1600)$ contributions must interfere
destructively in $M_{121}$. This consideration then leads to the extraction of
$M_{121}^{\Delta(1600)}$, which is shown in Fig.3, top, by its incoherent
contribution $\sigma_{121}^{\Delta(1600)}$ (thin dotted curve). Due to the
imposed destructive interference $\sigma_{121}^{\Delta(1600)}$ must be
a bit larger than $\sigma_{nn\pi^+\pi^+}$. The imposed destructive
interference is actually quite 
reasonable, if we consider the phase behavior of the two resonances. Both
$\Delta\Delta$ and $\Delta(1600)$ excitations reach their
poles in the region around $T_p \approx 1.4$ GeV. However,
whereas the resonance phase for $\Delta(1600)$ runs from zero to 180$^{\circ}$ 
as for usual resonances, the phase for the double resonance excitation
$\Delta\Delta$ runs from zero to 360$^{\circ}$. Hence, in the region of their
dominant contributions to $NN\pi\pi$, they are out of phase supporting thus the
imposed interference.

Having fixed now the issue of the unexpectedly large cross section in the
$nn\pi^+\pi^+$ channel, we focus next on the decomposition of $M_{101}$. Since
the pole of the Roper excitation is in the region of $T_p \approx 0.9$ GeV,
Roper and $\Delta\Delta$ run now essentially with the same resonance phase in
the region of interest, i.e. interfere constructively. By use of the
$\Delta\Delta$ cross section from the theoretical prediction we obtain then 
$\sigma_{101}^{N^*}$ as shown in Fig. 3, bottom, by the starred 
curve. Whereas  $\sigma_{101}^{N^*}$ rises up to $T_p \approx 0.9$ GeV as
expected by the theoretical prediction for the Roper excitation, its levels
off thereafter and decreases at still higher energies until it starts rising
again beyond 1.3 GeV, where the excitation of the $D_{13}$ resonance might be
expected. The essential observation here is that the Roper excitation
exhibits an energy dependence reminiscent of a Lorentzian shape as it is
adequate for a $NN^*$ s-channel resonance, but not for a $t$-channel associated
$N^*$ production. However, it can not be excluded at this stage that this
peculiar behavior is also possibly due to the interference of the Roper
excitation with that of higher-lying $N^*$ states.
If we compare our
solution for $\sigma_{101}^{N^*}$  in Fig. 3 with the theoretical prediction
for the $N^*$ excitation (dash-dotted curve Fig. 1, bottom),
which shows a continously rising cross section for the Roper excitation, then
we understand, why the theoretical 
 prediction fails so badly for the $pp\pi^0\pi^0$ channel beyond 0.9 GeV. 

Having understood $\sigma_{101}$ and $\sigma_{121}$ in some detail by the
contributions from $N^*$, $\Delta\Delta$ and $\Delta(1600)$ excitations, we
continue the discussion of $\sigma_{111}$ and extend it thereafter also to
$\sigma_{011}$. Since we have fixed the various resonance
contributions already by their confrontation with $\sigma_{101}$ and
$\sigma_{121}$ , we can calculate now $\sigma_{111}$ and $\sigma_{011}$ 
by use of the relations (3) - (6). The thus calculated contributions
$\sigma_{111}$ and $\sigma_{011}$ are shown in Fig. 4 by the dashed and
dot-dot-dot-dashed curves, respectively. Except of the highest energies
 $T_p >$ 1.3 GeV the calculated $\sigma_{111}$ is within the region of
 $\sigma_{111}(data)$ derived directly from the data.
The sum of all isospin contributions
is displayed by the solid curves. We see that the data in all channels are now
described quite reasonably. Even the $pn\pi^+\pi^0$ cross
section in the near-threshold region is reproduced somewhat better than by
the calculations of Ref. \cite{alv}, since it receives now also 
contributions from the $\Delta(1600)$ process. Still, since all isovector terms
have to vanish at threshold, even the inclusion of the $\Delta(1600)$  does
not contribute sufficiently at low energies, in order to achieve a satisfying
description of the data there, too. A solution of this problem may be found in
the $pn$ final state interaction (FSI) not taken into account so far. In our
procedure of isospin 
decomposition starting from the data in $nn\pi^+\pi^+$ and $pp\pi^0\pi^0$
channels we have effectively included the FSI there, since due to isospin
symmetry this FSI should be the same for $nn$ and $pp$ pairs apart from
Coulomb distorsions. However, in the
$pn$ system we have a much larger FSI, as borne out by the large $pn$
scattering length. Indeed, if we calculate the FSI effect on the total cross
section in a Migdal-Watson ansatz \cite{mig,wat} based on the 
experimental scattering lengths and effective ranges, we obtain an 
enhancement of the total cross section in the near-threshold region, which in
case of a $pn$ final state is roughly a factor of two larger than in case of
$pp$ and $nn$ final states. Thus inclusion of the $pn$ FSI effect leads to a
satisfactory understanding of the data also for the $pn\pi^+\pi^0$ channel.

In conclusion, new results for the total cross sections in $NN\pi\pi$ channels
have been presented. 
They reveal severe shortcomings in the previous theoretical 
descriptions of the two-pion production reaction. The isospin decomposition
based on all presently available total cross sections identify these
shortcomings to be due to the effect of a higher-lying $\Delta$ resonance --
most likely the $\Delta(1600)$ or possibly also the $\Delta(1700)$ -- not
considered in previous studies and due to the finding that the energy
dependence in the region of the Roper process exhibits a peculiar $s$-channel
like energy behavior.


We acknowledge valuable discussions with L. Alvarez-Ruso, V. Anisovich,
L. Dakhno, C. Hanhart, E. Oset and C. Wilkin on this issue.  
This work has been supported by BMBF
(06TU261), Forschungszentrum J\"ulich and  
DFG (Europ. Graduiertenkolleg 683). 

\end{document}